\documentclass[12pt,a4paper,final]{amsart}

\usepackage{graphicx}
\usepackage{amssymb}
\usepackage{amsmath}
\usepackage{pifont}
\usepackage[utf8]{inputenc}
\usepackage{pifont}
\usepackage[nolist]{acronym}
\usepackage{tabularx}
\usepackage{multirow}
\usepackage{floatrow}

\newfloatcommand{capbtabbox}{table}[][\FBwidth]

\usepackage{booktabs}
\usepackage{nccmath}

\usepackage{url}
\usepackage{subcaption}
\usepackage[dvipsnames]{xcolor}
\usepackage{algorithm2e}
\usepackage{fontawesome}
\usepackage{subcaption}
\usepackage{tikz}
\usepackage{listings}

\usepackage{xcolor}

\usepackage[misc,geometry]{ifsym} 

\definecolor{codegreen}{rgb}{0,0.6,0}
\definecolor{codegray}{rgb}{0.5,0.5,0.5}
\definecolor{codepurple}{rgb}{0.58,0,0.82}
\definecolor{backcolour}{rgb}{0.95,0.95,0.92}

\lstdefinestyle{mystyle}{
    backgroundcolor=\color{backcolour},   
    commentstyle=\color{codegreen},
    keywordstyle=\color{magenta},
    numberstyle=\tiny\color{codegray},
    stringstyle=\color{codepurple},
    basicstyle=\ttfamily\footnotesize,
    breakatwhitespace=false,         
    breaklines=true,                 
    captionpos=b,                    
    keepspaces=true,                 
    numbers=left,                    
    numbersep=5pt,                  
    showspaces=false,                
    showstringspaces=false,
    showtabs=false,                  
    tabsize=2
}

\usepackage{rotating}
\usepackage{pdflscape}
\usepackage{wasysym}
\usepackage{tablefootnote}

\usepackage{bookmark}

\usepackage{float}
\usepackage{hyperref}
\usepackage[noabbrev,sort&compress]{cleveref}

\newcommand{\cF}{\mathcal{F}}

\newcommand{\DeepUnHide}{\emph{DeepUnHide}~}

\begin{acronym}
    \acro{RS}{Recommender System}
    \acro{CF}{Collaborative Filtering}
    \acro{MF}{Matrix Factorization}
    \acro{MAE}{Mean Absolute Error}
    \acro{MSE}{Mean Squared Error}
    \acro{RMSE}{Root Mean Squared Error}
    \acro{CF4J}{Collaborative Filtering for Java}
    \acro{BeMF}{Bernoulli Matrix Factorization}
    \acro{RPI}{Reliability Prediction Index}
    \acro{KNN}{K-Nearest Neighbours}
    \acro{PMF}{Probabilistic Matrix Factorization}
    \acro{NMF}{non-Negative Matrix Factorization}
    \acro{BNMF}{Bayesian non-Negative Matrix Factorization}
    \acro{URP}{User Ratings Profile Model}
    \acro{NN}{Neural Network}
    \acro{DL}{Deep Learning}
    \acro{ML}{Machine Learning}
    \acro{CNN}{Convolutional Neural Network}
    \acro{MLN}{Multilayer Neural Network}
\end{acronym}

\author[]{Jesús Bobadilla}
\address{ETSI de Sistemas Inform\'aticos, Universidad Polit\'ecnica de Madrid. C.\ Alan Turing s/n, 28031. Madrid, Spain.}
\email{jesus.bobadilla@upm.es}
\author[]{\'Angel Gonz\'alez-Prieto}
\address{ETSI de Sistemas Inform\'aticos, Universidad Polit\'ecnica de Madrid. C.\ Alan Turing s/n, 28031. Madrid, Spain.}
\email{angel.gonzalez.prieto@upm.es}
\author[]{Fernando Ortega}
\address{ETSI de Sistemas Inform\'aticos, Universidad Polit\'ecnica de Madrid. C.\ Alan Turing s/n, 28031. Madrid, Spain.}
\email{fernando.ortega@upm.es}
\author[]{Ra\'ul Lara-Cabrera}
\address{ETSI de Sistemas Inform\'aticos, Universidad Polit\'ecnica de Madrid. C.\ Alan Turing s/n, 28031. Madrid, Spain.}
\email{raul.lara@upm.es}

\begin{document}

\newtheorem{thm}{Theorem}[section]
\newtheorem{prop}[thm]{Proposition}
\newtheorem{lem}[thm]{Lemma}
\newtheorem{cor}[thm]{Corollary}

\newtheorem{defn}[thm]{Definition}
\newtheorem{as}{Assumption}

\newtheorem{rmk}[thm]{Remark}
\newtheorem{ex}[thm]{Example}

\usetikzlibrary{bayesnet}
\usetikzlibrary{arrows}

\title[Deep Learning feature selection to unhide demographic factors]{Deep Learning feature selection to unhide demographic recommender systems factors}

\date{}

    \begin{abstract}
    Extracting demographic features from hidden factors is an innovative concept that provides multiple and relevant applications. The matrix factorization model generates factors which do not incorporate semantic knowledge. This paper provides a deep learning-based method: \textit{DeepUnHide}, able to extract demographic information from the users and items factors in collaborative filtering recommender systems. The core of the proposed method is the gradient-based localization used in the image processing literature to highlight the representative areas of each classification class. Validation experiments make use of two public datasets and current baselines. Results show the superiority of \textit{DeepUnHide} to make feature selection and demographic classification, compared to the state of art of feature selection methods. Relevant and direct applications include recommendations explanation, fairness in collaborative filtering and recommendation to groups of users.\newline{}
    \null\hspace{0.5cm}\textbf{Keywords:} Feature selection, collaborative filtering, demographic information, matrix factorization, gradient based localization, deep learning.
    \end{abstract}

\maketitle
    


\section{Introduction}\label{sec:introduction}
\sloppypar{\ac{RS}~\cite{Madadipouya2017Jul,Saquib2017} are playing an important role in our society: they provide useful information to the users by recommending highly demanded products and services. Remarkable examples of \ac{RS} are: Amazon, Netflix, TripAdvisor and Spotify. \ac{RS} are implemented by means of several filtering strategies, mainly the collaborative~\cite{Madadipouya2017Jul,Saquib2017}, content~\cite{Zamani2018Oct}, demographic~\cite{AlShamri2016May}, context~\cite{Villegas2018Jan} and social~\cite{Rezvanian2019} ones. Most of the commercial \ac{RS} are based on hybrid models that combine \ac{CF} with some other filtering approaches. In the early ages of \ac{RS} research, \ac{CF} was implemented using the \ac{KNN} algorithm~\cite{Bobadilla2013Jul}: it is easy to understand, to implement and to analyse, since it can be considered as a white-box method. This approach has also been updated and improved in the recent years with promising approaches like hybrid methods \cite{Barragans-Martinez2010Nov} or adding information theoretic quality measures \cite{Jiang2019Dec}. Nevertheless, the \ac{KNN} main drawbacks are its lack of scalability and its poor accuracy.

Due to the exposed \ac{KNN} drawbacks, this memory-based algorithm has been replaced by model-based ones, mainly the \ac{PMF}~\cite{mnih2008probabilistic} and its variations and improvements, such as the \ac{NMF}~\cite{NIPS2000_1861}, Bayesian NMF (BNMF)~\cite{Hernando2016Apr} of two-level MF (TLMF)~\cite{Li2016Nov}. Currently, research is also focusing on \ac{DL}~\cite{Mu2018Nov,Bobadilla:2020,Huang2020Jun} based approaches. Model-based approaches are scalable and accurate, but they act as a black box, making it difficult to address some \ac{RS} goals such as recommendations explanation or improvements in the beyond accuracy goals such as fairness, diversity, reliability or serendipity. The explained research evolution on \ac{RS} is relevant to this paper, since it makes use of an architecture that combines the \ac{MF} and \ac{DL}  approaches, trying to unhide the \ac{MF}  black box model.}

\begin{figure}[!h]
    \centering
    \includegraphics[width=0.7\columnwidth]{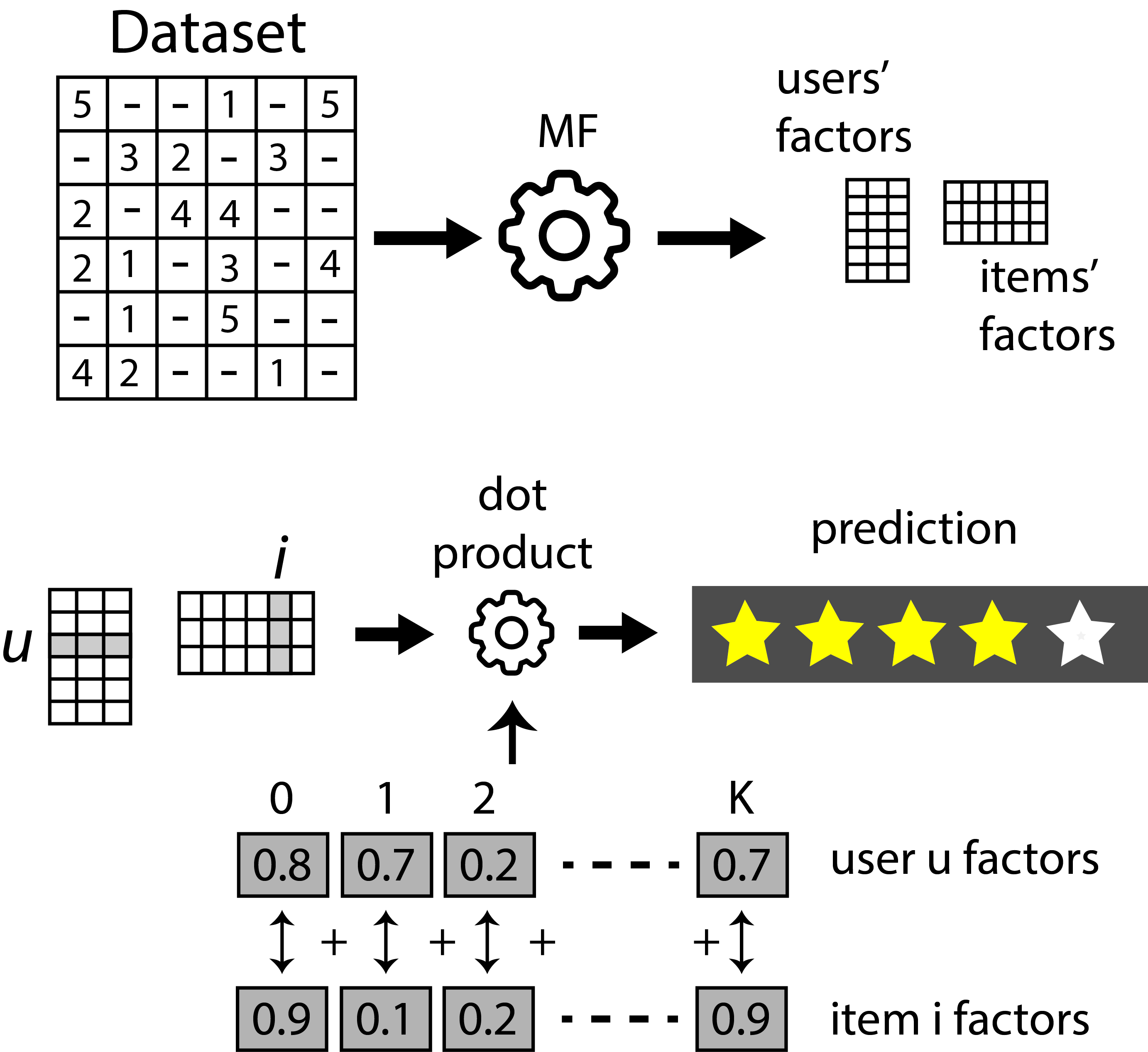}
    \caption{Matrix factorization and the dot product to make predictions}\label{fig:1}
\end{figure}

\ac{CF} \ac{RS} datasets are really sparse~\cite{Bobadilla2020}, and \ac{MF} models make a reduction of dimensionality to obtain compressed and dense versions of them. In the \ac{MF} models, each user is represented by a reduced number of $K$ factors (real numbers) that encode the user’s essence. Each dataset item is represented in the same way. \Cref{fig:1} shows the \ac{MF} basic operational; on its top the compression is represented, where a sparse matrix of ratings is converted in two dense matrices of factors: the users' and the items' ones. To predict the rating of a user to an item, the dot product is used; recommendations to each user are just those items with the best predictions. The bottom of \cref{fig:1} shows the essence of the dot product: predictions will be high (the user will like the item) when the users’ factors and the item’s factors are significant, and they also match (they have similar values). In the \cref{fig:1} example, we can observe that first and the last factors have similar values: they match each other. The user and the item third factors match, but they are not relevant. Finally, the second factors do not match. Each user's factor encodes some features combination; a simplistic view could state that the fifth factor encodes: the user is female and young, whereas third factor encodes that she is a female and she likes musical films. Please note that each feature can be coded in several factors. Items are also encoded in $K$ factors; e.g.: Avatar film factor fourth could encode `young', `scify' and `popular', whereas the first factor could encode `scifi' and `current'. 

\begin{figure*}[!h]
    \centering
    \includegraphics[width=0.85\textwidth]{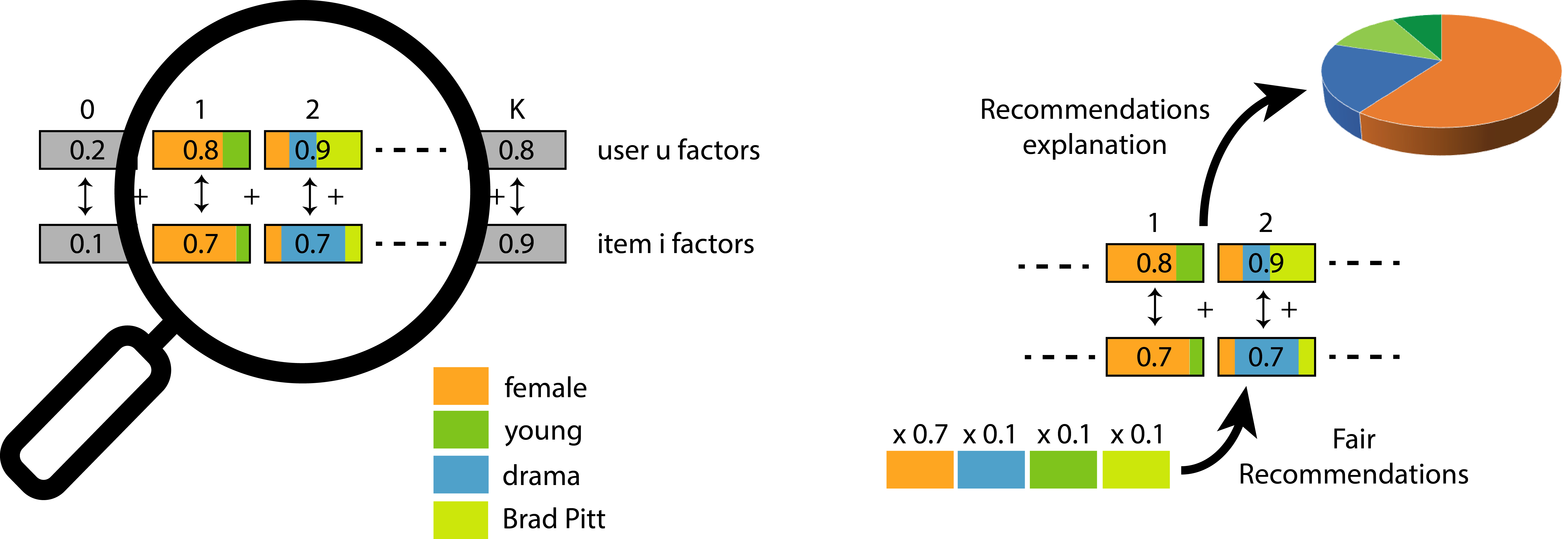}
    \caption{Matrix Factorization hidden factors semantic and applications}\label{fig:2}
\end{figure*}

It would be great if the \ac{MF} models could return the semantic meaning of each one of the $K$ chosen factors, but in fact none of the \ac{MF} models can do it. \ac{MF} can predict how much a user will like a not voted item, and even it can relate users or items by measuring the distance between their factors, but \ac{MF} cannot directly establish why it predicts that you will like an item or you will not like another one. \ac{MF} just acts on the ratings; it does not directly process demographic information (gender, age, etc.) because it has not been designed to do it. To better understanding this concept \cref{fig:2} shows, in grey color, the hidden factors, both the items and the users; it means that we do not know the semantics of the hidden factors. It would need some algorithmic process to show us all the \ac{MF} factors from a different perspective, in the same way that infrared cameras show our environment. This algorithmic process is represented in \cref{fig:2} as a magnifier glass. The proposed method in this paper performs the represented magnifier glass function, and it can show the semantic meaning of \ac{MF} factors in those features we have selected (usually demographic ones).

Inside the magnifier it is shown a coloured new information. It tells us about the different degrees of demographic features each factor encodes; e.g.: the user u factor 1 has a big proportion of female feature, followed by a less proportion of young user; in the same way, item i factor 2 mostly encodes a drama film that could like to Brad Pitt's female fans. To get this additional information is important, since it opens the door to design improved methods in different \ac{RS} research fields, such as recommendation to groups of users or recommendation to users who share minority preferences. In the former case a representative virtual user can be obtained by combining the factors values with the factor demographic proportions, it the latter case a new feature `minority' can be created to identify minority users. Beyond the two previous examples, we have selected two main \ac{RS} research fields where the proposed method can be particularly important: explanation of recommendations and fair recommendations. The right side of \cref{fig:2} shows the proposed strategies to address both objectives: recommendations explanation can be based on the impact that each demographic feature has in the recommendation; in this example we can inform to the user that the recommendation is mostly based on the feminine component of the film, and to a much lesser extent because it is a drama film that usually like Brad Pitt's female fans. Fair recommendations can be obtained acting on the dot product stage, by positively weighting the desired demographic features; \cref{fig:2} shows an example where female factors are privileged with most of the prediction importance: it is an example of fair recommendations applied to the female minority group (most of the \ac{RS} datasets are biased in gender and age).

The recommendation fairness approach is innovative in the field. Some of the \ac{CF} fairness research has been focused on the \ac{KNN} algorithm, since it is a white box approach that allows to design tailored solutions, such in~\cite{Burke} where fairness is obtained by choosing balanced neighborhoods. In the model-based \ac{CF} data biases has been studied as a source of unfair recommendations~\cite{Ekstrand2020}. Diversity in recommendations leads to unfair results and discrimination~\cite{Leonhardt2020}, but it is also necessary to balance different goals such as fairness, accuracy, diversity and novelty. Bias disparity has been defined as ``how much an individual's recommendation list deviates from his or her original preferences in the training set''~\cite{Tsintzou2018Nov}. Research in \ac{CF} fairness has been focused on study the datasets bias rather than to design models to tackle the problem: ``teams typically look to their training datasets, not their machine learning models, as the most important place to intervene to improve fairness in their products''~\cite{Holstein:2019}. A review of the \ac{RS} fairness issue is presented in~\cite{Chouldechova:2020}, where some frontiers in the field are highlighted. The \ac{MF} method cannot easily manage the two main sources of imbalanced data: observation bias and population imbalance~\cite{Yao2017}. As it can be seen, no model-based \ac{CF} approach have been conducted in the same line as the proposed one in this paper; additionally, our method makes use of \ac{DL}  technology, and this is a specific field with little research made in the \ac{RS} fairness issue: in the \ac{DL}  based \ac{RS} survey~\cite{Mu2018Nov} fairness is not mentioned, not even in its research directions section. In the review paper~\cite{Batmaz2019Jun} fairness is not addressed, either.

\sloppypar{The recommendation explanations~\cite{Nunes2017Dec} research field has a \ac{KNN} based area~\cite{Zanker2020,Papadimitriou2012May} that is not relevant to our model-based approach. Several strategies have been designed to address \ac{CF} explanations: graphs have been used to relate recommendations sources~\cite{Lully2018Jan}, explanations to group recommendations are also designed based on the group social reality looking for positive reactions from the members of each group~\cite{Quijano-Sanchez2017Jun}.}

Recommendations have also been explained by using temporal information of the ratings~\cite{Bharadhwaj2018Nov,Valdiviezo-Diaz2019Aug}. Trees have been shown where neighbour users and related items are drawn around the recommended user position~\cite{Hernando2013Aug}. As far as we know there is not a published \ac{DL}  model to address the explanations of \ac{CF} recommendations made through \ac{MF} factors; nevertheless, there is a paper that emphasizes the importance of the demographic information versus content information in \ac{CF} explanations~\cite{bilgic2005explaining}. 

\begin{figure}[!h]
    \centering
    \includegraphics[width=0.7\columnwidth]{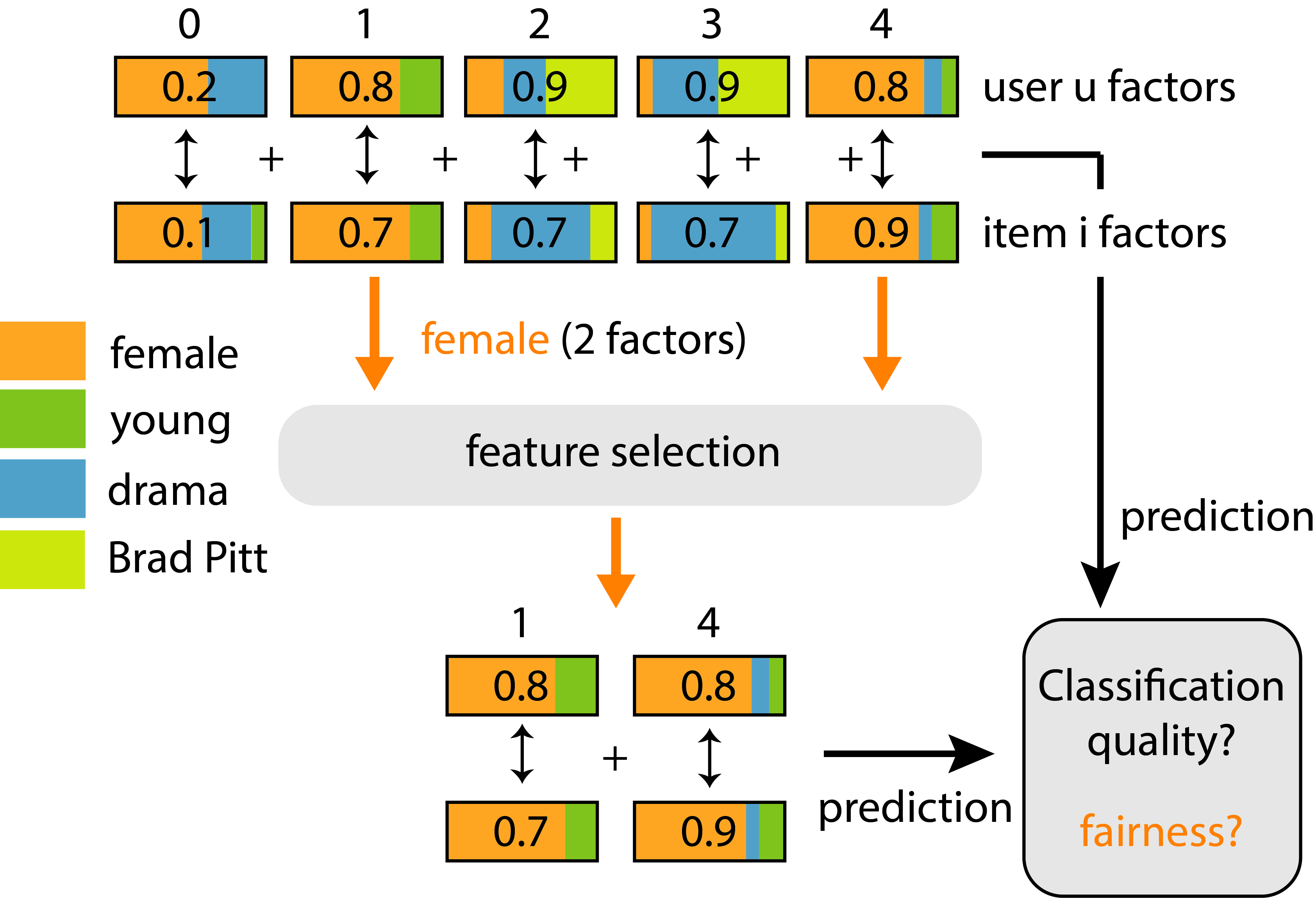}
    \caption{Feature selection based on demographic information}\label{fig:3}
\end{figure}

\sloppypar{Feature selection is also related to this paper, since we test the proposed method results by selecting the most promising factors, discarding the rest and measuring the impact of the filtering. \Cref{fig:3} shows the concept: we can compare the classification results on a demographic target (e.g: gender) by using the $K$ existing factors versus the classification results just using a subset of the $K$ factors (only 2 factors in this example). The more similar the classification accuracy, the best the performed feature selection. To claim the superiority of the \ac{DL}  proposed feature selection method we compare it with a set of popular feature selection baselines: logistic~\cite{Ng2020}, entropy~\cite{Jiang2015Jul}, variance~\cite{Wang2018Aug} and Principal Component Analysis (PCA)~\cite{Jolliffe2002}. Fairness impact can also be tested, since focusing on demographic feature selection we can obtain demographic-based fair recommendations.}

\begin{figure}[!h]
    \centering
    \includegraphics[width=0.9\columnwidth]{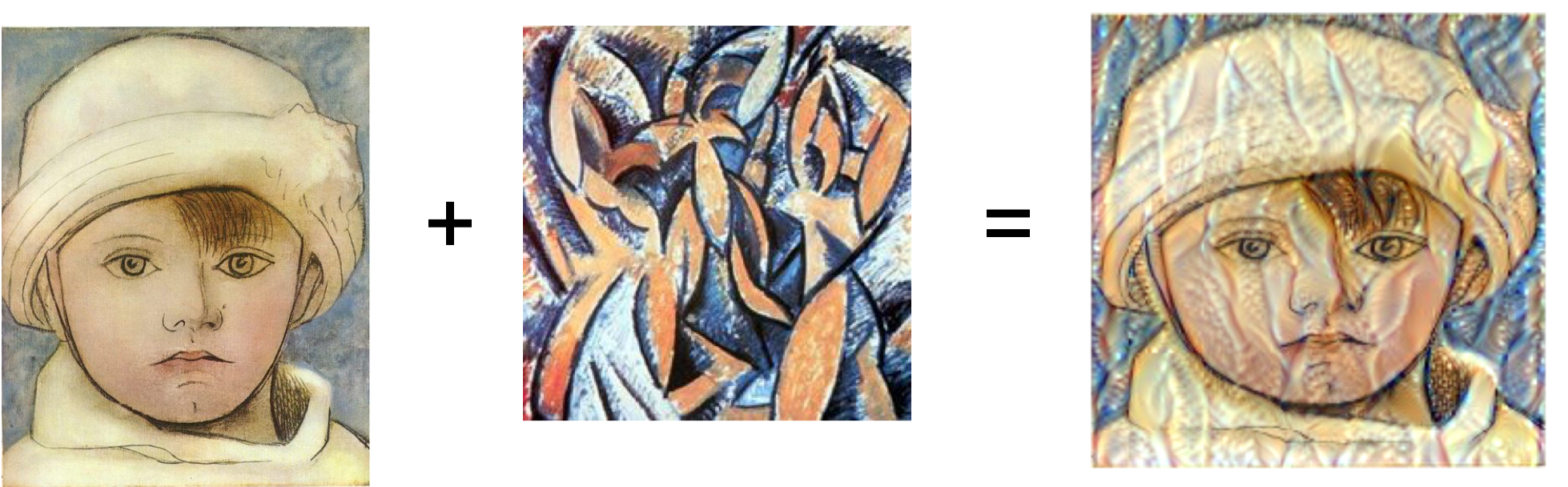}
    \caption{Example of Neural Style Transfer}\label{fig:4}
\end{figure}

Finally, there is a \ac{DL}  research field we have borrowed from the image processing area to act as a kernel of the proposed method: the deep networks gradient-based localization. Grad-Cam~\cite{Selvaraju2017Oct} uses any target concept (say `cat') in a classification network to generate a localization map. It active the relevant areas, in the image, that encode the concept. Grad-Cam generalizes the CAM research~\cite{Zhou2016Jun}, where generic localizable \ac{DL}  representations are built. CAM uses the global average pooling as a structural regularizer~\cite{Lin2014NetworkIN}. Neural style transfer (NST)~\cite{Jing:2019} is also a reference to the proposed approach; in this case, a source image is converted to the style of another image that acts as a target. This is made by minimizing the gradient between the source image and one or several chosen filters of a Convolutional Neural Network (CNN). Our method performs this operation, using a noisy source instead of a regular image. The NST was introduced by~\cite{Gatys:2016} using intermediate layers of the VGG-19~\cite{Wen:2019} network to catch different styles. The style representation has been based on the Gram matrix~\cite{NIPS2015_5633} by matching style and stylised images. To graphically show the concept, we have designed an NST and fed it with two Picasso's pieces of art; as it can be seen in \cref{fig:4}, the style image has been passed to the source image. We have chosen the 'block1\_pool' and 'block2\_pool' layers of the VGG19 network as style filters.

The rest of the paper has been structured as follows: in \Cref{sec:model} the proposed method is explained, and the experiments design is defined. \Cref{sec:experiments} shows the experiments' results and their discussions. Finally, \Cref{sec:conclusions} contains the main conclusions of the paper and the future works.

\section{Model}\label{sec:model}

\begin{figure}[!h]
    \centering
    \includegraphics[width=0.95\columnwidth]{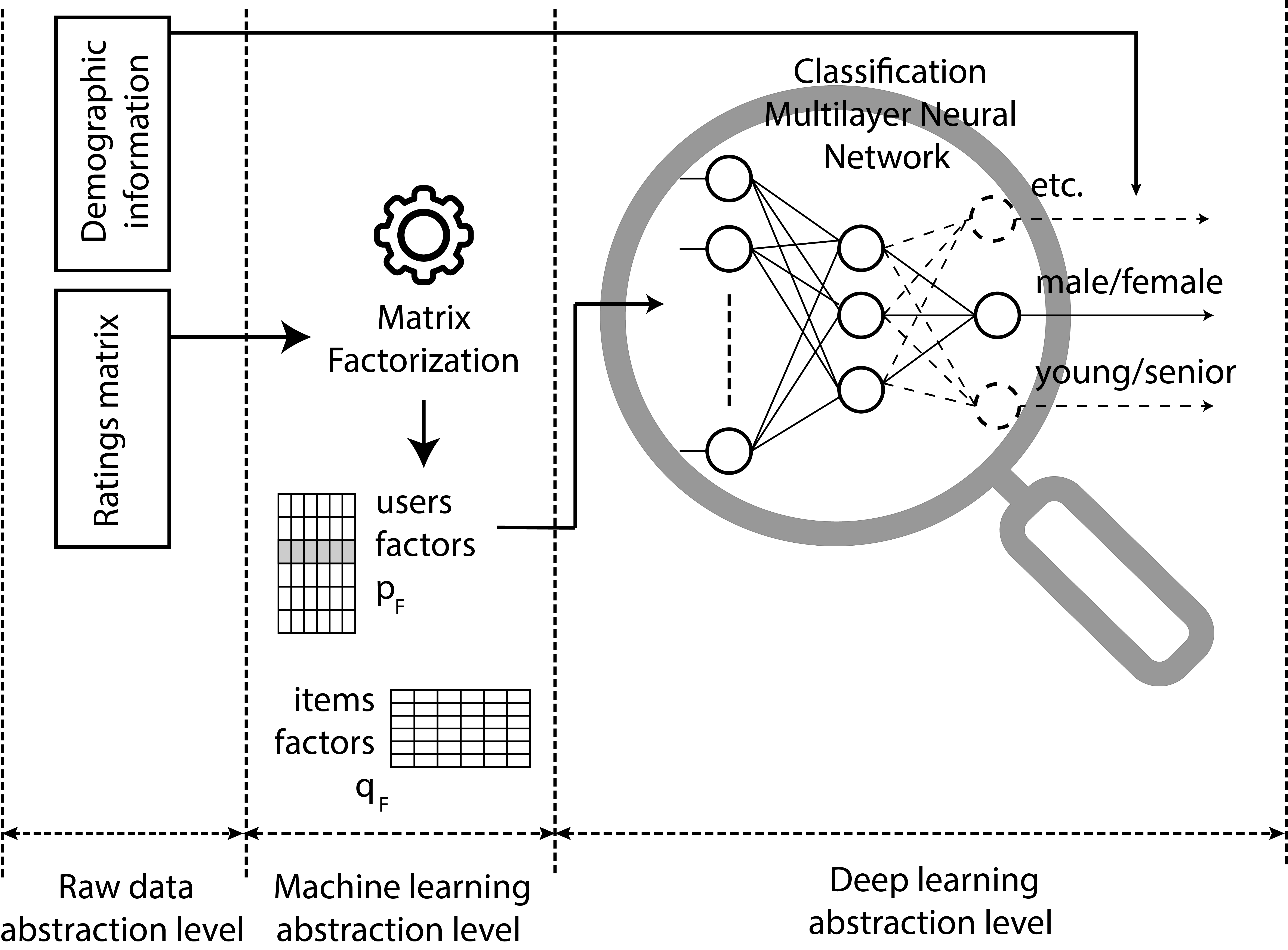}
    \caption{\DeepUnHide Architecture}\label{fig:5}
\end{figure}

The proposed method to unhide \ac{MF} factors is inspired in the gradient-based localization~\cite{Gatys:2016} and the neural transfer learning~\cite{Wen:2019} techniques. Since it is a \ac{DL}  approach to unhide factors, we have called it \DeepUnHide. First, we must design an architecture and then to apply the proposed method to it. \Cref{fig:5} shows the \DeepUnHide architecture; it is composed of three abstraction levels: raw data, \ac{ML} and \ac{DL}. The raw data abstraction level feeds the architecture with the necessary information, in our case it just needs the \ac{CF} matrix of ratings and the selected demographic information (gender, age, etc.). The \ac{ML} abstraction level is in charge of providing the \ac{MF} hidden factors. For that purpose, in the proposed model we used standard \ac{MF} methods as \ac{PMF}~\cite{mnih2008probabilistic}. 

As it can be seen in \cref{fig:5}, we will only take the users’ factors, since we want to explain recommendations based on demographic data related with users. It is also possible to explain recommendations based on demographic data related to items (genre, popularity, director, etc.), in this case we would take the items' factors to feed the following architectural layer. Our last abstraction level is the \ac{DL}  one: we make use of a \ac{MLN} to classify users by demographic information. In this paper, as an example, we have chosen the male/female and young/senior groups. Notice that our objective is not to classify users: We train this \ac{MLN} to feed the proposed method with the learned weights of the neural network. 

\subsection{Gradient localization for image processing}

\begin{figure}[!h]
    \centering
    \includegraphics[width=0.7\columnwidth]{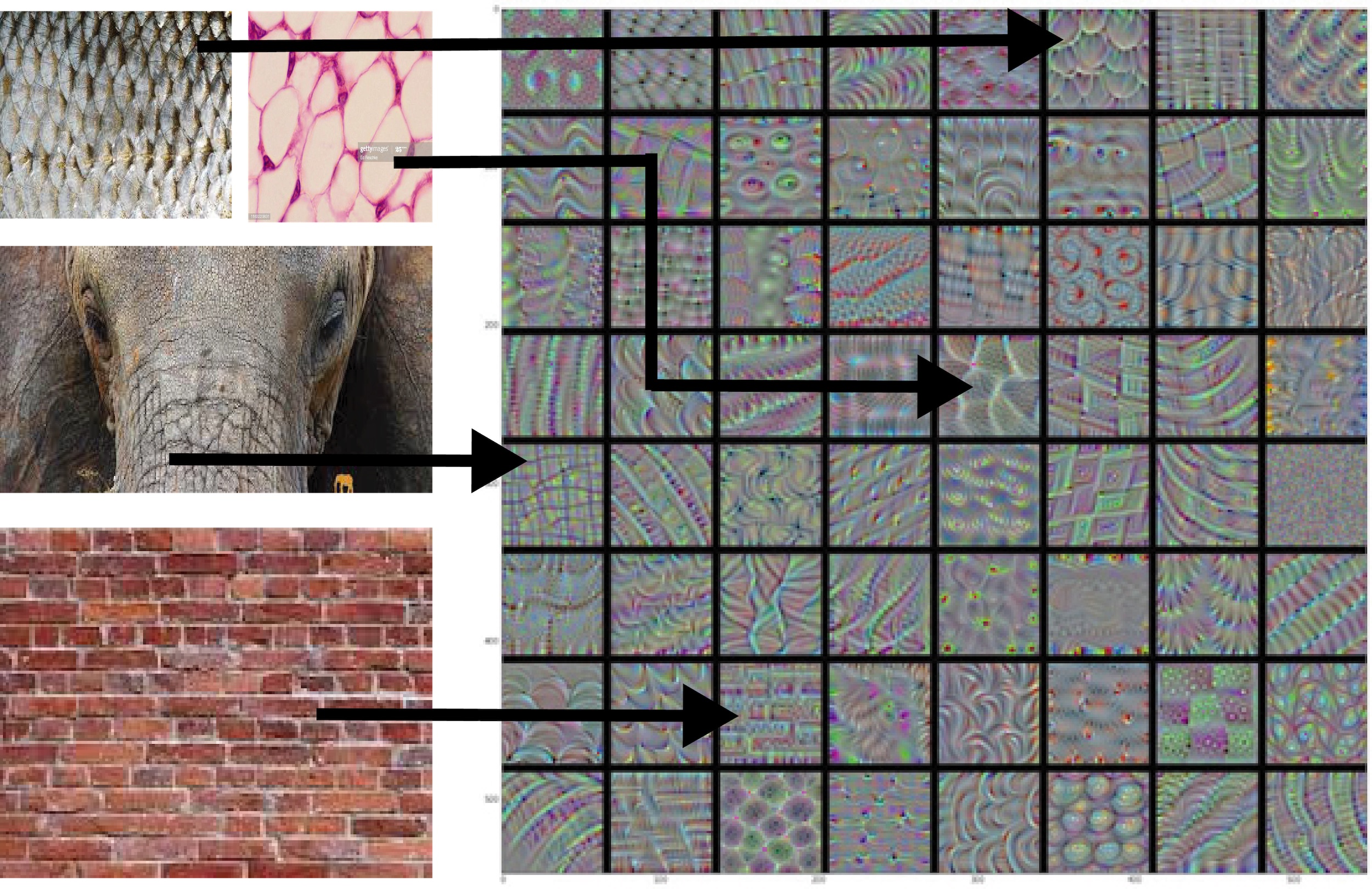}
    \caption{Learnt patterns in the 64 filters of the VGG16 Block4\_conv1 layer}\label{fig:6}
\end{figure}

Once the architecture is set, we can explain the details of the proposed \DeepUnHide method, in which we will use the \ac{MLN} information shown with the magnifier glass metaphor in \cref{fig:5}. In the gradient based localization~\cite{Gatys:2016}, conceptually we make a process similar to the one shown in \cref{fig:4}, but there are some key differences; one of them is that we do not have a defined source image: we will use a noise source. Another difference is that instead of using images, our source is a list of hidden factors. To graphically explain the first concept, we make use of the image processing field: \Cref{fig:6} visually shows the learnt pattern of each filter in the VGG16 Block4\_conv1 layer. These filters help to classify some of the images used to train the VGG16 \ac{CNN}~\cite{KrishnaswamyRangarajan2020Feb}. Each one of these map activations can serve to get the input pattern that best active the corresponding filter: this is the gradient-based localization key. To obtain the mentioned input patterns we apply an initial random noise image to the input of the classification neural network. Afterwards, we make use of the gradient descent algorithm to iteratively change the input image until the loss function is minimized. Here, the loss function is the distance between each activation map the input image generates and the chosen filter values.

\begin{figure}[!h]
    \centering
    \includegraphics[width=0.75\columnwidth]{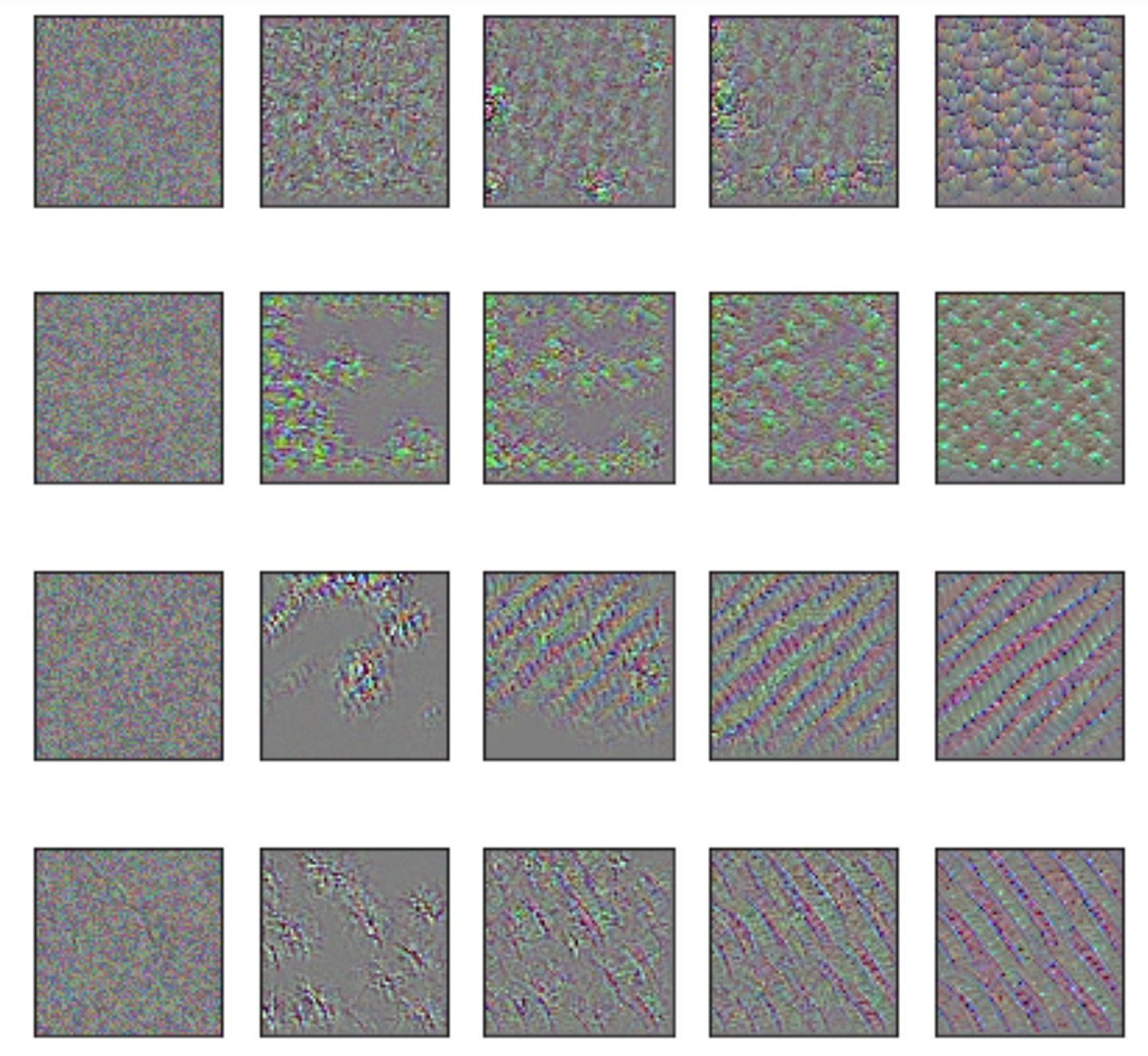}
    \caption{Gradient descent intermediate images obtained from a noisy picture to four of the activation maps in \cref{fig:6}}\label{fig:7}
\end{figure}

Each row in \cref{fig:7} shows an example of the aforementioned process: rows contain the gradient descent result of applying a noisy image (left) to the \ac{CNN}, by minimizing the loss differences with four of the filters in \cref{fig:6}. Each of the right-most images of \cref{fig:7} shows the input pattern that maximizes its corresponding filter detection. They can be considered as representative patterns in some areas of different types of images. What is relevant to us is the concept that using gradient descent on a pre-trained \ac{MLN} we can find representative input patterns of the output targets. Moving to the \ac{RS} field and using the \DeepUnHide Architecture (\Cref{fig:5}) we can find representative patterns of demographic features; more precisely: we can find the user factors values that best represent the male, female, young, senior, etc. users.

\subsection{Gradient localization \DeepUnHide}

\begin{figure}[!h]
    \centering
    \includegraphics[width=0.55\columnwidth]{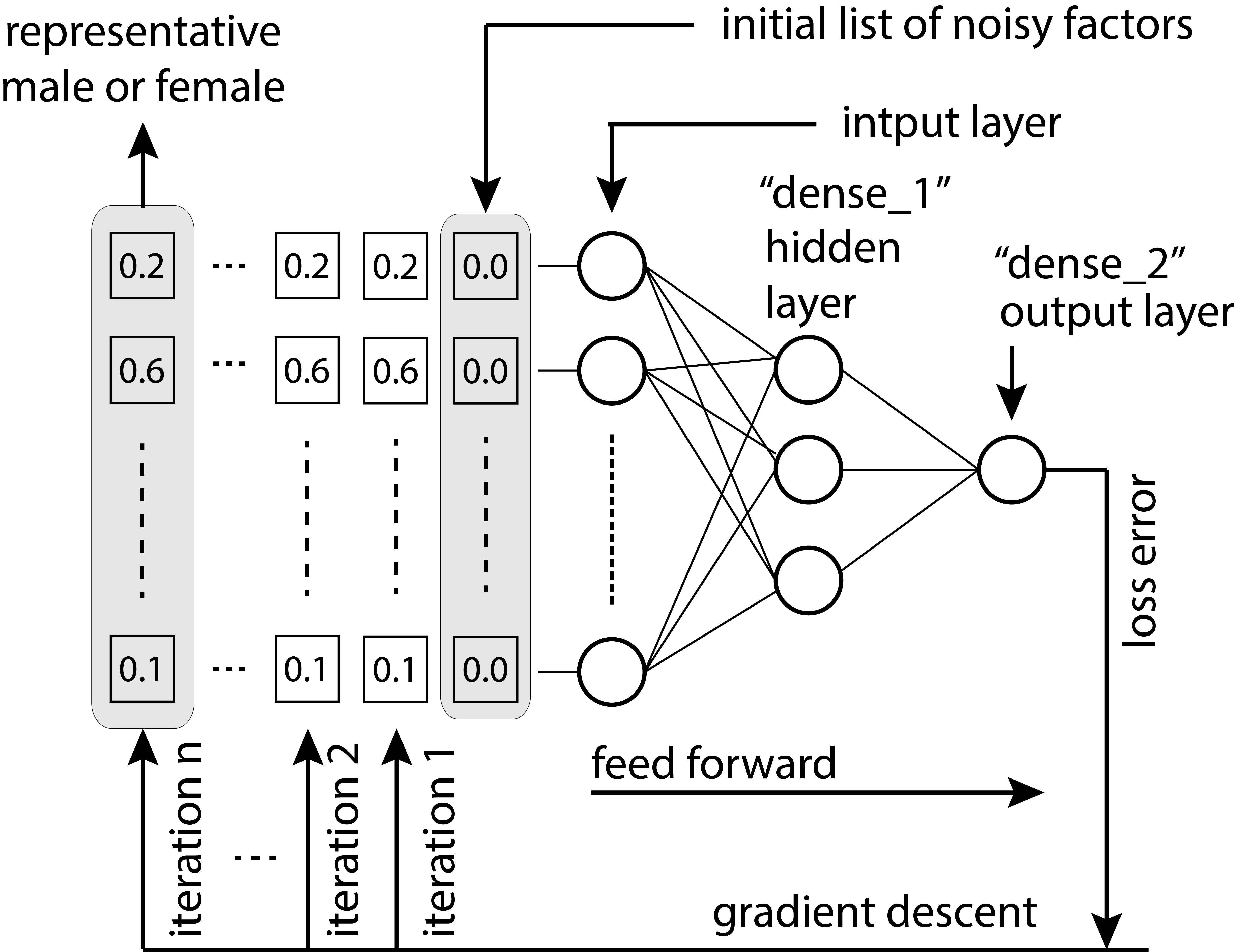}
    \caption{\DeepUnHide gradient based localization method}\label{fig:8}
\end{figure}

The proposed \DeepUnHide method is explained in \cref{fig:8}. Starting from the trained \ac{MLN} in the \DeepUnHide architecture (\Cref{fig:5}), an initial random list of factors, or an initial list of factors filled to 0 are presented to the \ac{MLN} (``initial list of noisy factors''). Using this starting vector, a feed forward process is conducted to obtain the prediction; then an output and a loss error are obtained; e.g.: we expect the 0 value in the female case or the 1 value in the male one. Then, the gradient descent algorithm obtains the input values that minimize the error, in the first iteration. As usual in the gradient descent operative, the process is repeated until the error reaches a threshold or until a prefixed number of iterations have been run. At the end of the process we get the factors values of the representative demographic user (male, female, young, etc.). Please note that repeating this process for a set of demographic features we can obtain the proportions shown in \cref{fig:2,fig:3}; it demographically unhides the users’ factors. This process can also be done to demographically unhide the items’ factors. 

\subsection{Mathematical formulation of \DeepUnHide}\label{sec:math-formulation}

To be precise, suppose that we have an initial spare matrix of ratings $R = (r_{u,i})$, where $r_{u,i}$ is the rating that user $u$ assigned to item $i$ (say, in a discrete scale from $0$ to $N$). Denote the number of users in the model by $U$ and the number of items by $I$ so that $R$ is a $U \times I$ matrix.

The objective of \ac{PMF} is to find a dense matrix $\tilde{R}$ that coincides with $R$ as much as possible in the known ratings. For this purpose, we look for a factorization of the form $\tilde{R}= PQ^t$ where $P$ is a $U \times K$ matrix and $Q$ is a $I \times K$ matrix. The interpretation of these matrices is that the $u$-th row of $P$, $p_u$, is the $K$-dimensional vector of hidden factors of the user $u$; and analogously for the $i$-th row of $Q$, $q_i$, for the hidden factors of the item $i$. In this way, we want to minimize the cost function
$$
    ||R - \tilde{R}||_2^2 = ||R - PQ||_2^2 = \sum_{r_{u,i} \neq \bullet} (r_{u,i}-p_u \cdot q_i)^2,
$$
where $r_{u,i} \neq \bullet$ denotes that the rating of the user $u$ to the item $i$ is known, and $||\cdot||_2$ denotes the usual euclidean distance between the vectors of known ratings. A standard gradient descent algorithm with regularization for minimizing the cost function leads to the update rule

\begin{align*}
    p_u & \leftarrow p_u + \gamma \left(\sum_{r_{u,i} \neq \bullet} (r_{u,i} - p_u \cdot q_i) q_i - \lambda p_u\right), \\
    q_i & \leftarrow q_i + \gamma \left(\sum_{r_{u,i} \neq \bullet} (r_{u,i} - p_u \cdot q_i) p_u - \lambda q_i\right).
\end{align*}

\sloppypar{Here, $\gamma, \lambda > 0$ are two hyperparameters of the training method (the steps of the gradient descent).}

As mentioned above, in this paper we will focus on unhiding the users factors, $p_u$, so we pull apart the items factors $q_i$. Now, we focus on some demographic binary classification into majority/minority group (say male/female or young/senior). For that purpose, we consider a \ac{MLN},
$$
    h: \mathbb{R}^K \to \mathbb{R},
$$
trying to fit the perfect classification given by $h(p_u) = 1$ if $u$ belongs to the majority group and $h(p_u) = 0$ if $u$ belongs to the minority group. The neural network $h$ is trained with the usual gradient descent optimization on its parameters (the so-called backpropagation method).

Once this \ac{DL}  step is completed, we look for the factors that maximize the expectancy of $h$ of predicting a given demographic group. Hence, we fix an objective target $t = 1,0$ ($t =1$ if we are focusing on the majority group and $t = 0$ if we are interested in the minority group). Now, we define the cost function
$$
    \cF_t: \mathbb{R}^K \to \mathbb{R}_{\geq 0}, \quad \cF_t(p) = \frac{1}{2}(t-h(p))^2.
$$
That is, $\cF_t(p) = 0$ if and only if $h(p) = t$, which means that $p \in \mathbb{R}^K$ is the `archetypal' user factors of a member of the demographic group $t$. In order to minimize $\cF_t$, we use a standard gradient descent algorithm. For this purpose, observe that the gradient of $\cF_t$ is given by
$$
    \nabla \cF_t(p) = -(t-h(p)) \,\nabla h(p).
$$
Observe that the gradient $\nabla h(p)$ can be easily computed in terms of the internal weights of the \ac{MLN} by means of the usual backpropagation method. Therefore, the usual gradient descent method leads to the update rule
$$
    p \leftarrow p + \eta (t-h(p))\,\nabla h(p).
$$
Here, $\eta>0$ is a hyperparameter of the training process that corresponds to the step of the gradient descent. The initial guess for $p$ can be taken as a random vector drawn from a uniform distribution, or just as the zero vector. This process is the so-called gradient localization in the image processing literature.

As a result of this optimization step, we get two preferred user factors $p_M, p_m \in \mathbb{R}^K$ for the majority and the minority group. As mentioned above, these can be understood as the factors of a representative user of each demographic group. Let us we write the components of these vectors as $p_M = (p_M^1, p_M^2,\ldots, p_M^K)$ and $p_m = (p_m^1, p_m^2,\ldots, p_m^K)$. In order to interpret these vectors as amount of affinity, we normalize them to take values in the interval $[0,1]$ as
\begin{align*}
    \tilde{p}_M & = \frac{1}{p_M^{\textrm{max}} - p_M^{\textrm{min}}} \left(p_M^1 - p_M^{\textrm{min}}, \ldots, p_M^K - p_M^{\textrm min} \right), \\
    \tilde{p}_m & = \frac{1}{p_m^{\textrm{max}} - p_m^{\textrm{min}}} \left(p_m^1 - p_m^{\textrm{min}}, \ldots, p_m^K - p_m^{\textrm min} \right),
\end{align*}
where $p_M^{\textrm{max}} = \max_j p_M^j$, $p_M^{\textrm{min}} = \min_j p_M^j$, $p_m^{\textrm{max}} = \max_j p_m^j$ and $p_m^{\textrm{min}} = \min_j p_m^j$. In this way, a value of $\tilde{p}_M^j$ (resp.\ $\tilde{p}_m^j$) near to $1$ shows that the $j$-th factor characterizes a hidden characteristic that is like-minded to the majority (resp.\ minority) group whereas a value near to $0$ evidences that the $j$-th factor measures a characteristic that is typically disliked by the majority (resp.\ minority) group.

This idea leads to a feature selection criterion of relevant factors for the majority (resp.\ minority) group by sorting the factors $j = 1, \ldots, K$ by decreasing value of $\tilde{p}_M^j$ (resp.\ $\tilde{p}_m^j$). In this way, fixed a number of desired factors $N < K$, we can obtain the subsets $K_M$ and $K_m$ of the most relevant $N$ factors for the majority and minority group, respectively.

Moreover, this information can also be used for proving an absolute measure of the importance of each factor to the dichotomy majority/minority, as the distance of this factor between the majority archetypal user and the minority archetypal user. Hence, we take
$$
    \textrm{relevancy}(j) = |\tilde{p}_M^j-\tilde{p}_m^j|.
$$
In this way, high values of $\textrm{relevancy}(j)$ evidences that the $j$-th factor has typically a large variation from a demographic group to another (say, it is high in the majority group and low in the minority group, or vice-versa), whereas low values of $\textrm{relevancy}(j)$ point out that this factor is similar in both demographic groups. Therefore, factors with high relevancy are the best indicators of the membership of an user to a group. Again, this relevancy can also be used as a feature selection criterion for choosing the factors that are more relevant for the associated classification problem.

\subsection{Implementation of the model}

Algorithm 1 implements the \DeepUnHide internals by using Keras and Python. Since it is a really short piece of code it has been considered useful to include the algorithm in this paper in order to explain the method, to easily reproduce the experiments and to base some future works on it. Previous to running the shown procedure, we have trained a \ac{MLN} using Keras. In our example, we have chosen an architecture with 2 layers, being 'dense\_1' the hidden layer and 'dense\_2' the output layer (see \cref{fig:8}). Line 2 establishes the output of the model: ``dense\_2'' layer, in the neural network drawn in \cref{fig:8}. Line 3 sets the loss function: in our case, the \ac{MLN} correctly predicts demographic features. Line 4 makes the hard work, obtaining the gradients of the input with regard to the loss. Line 5 just normalizes the gradients. Line 6 returns the loss and the gradient obtained from the input (input factors). These input factors are initialized in line 7. Then, a gradient descent loop is set, in line 8, to run each established iteration and to obtain the new gradient values, in line 9. Finally, the input factors are updated in little 0.1 steps, in line 10. This line of code generates the subsequent input factors shown on the left of \cref{fig:8}, where the result is shown in grey background.

\begin{lstlisting}[language=Python]
def factors(gender):
    output = model.get_layer("dense_2").output
    loss = 1/2*(gender-output)**2
    gradient = K.gradients(loss, model.input)[0]
    gradient /= (K.sqrt(K.mean(K.square(gradient))) + 1e-5)
    iteration = K.function([model.input], [loss,gradient])
    input_factors = np.expand_dims(np.zeros(userFactors.shape[1]), axis=0)
    
    for i in range(20):
        loss_value, gradient_value = iteration([input_factors])
        input_factors += gradient_value * 0.1
    return input_factors

    MALE, FEMALE = 1., 0.
    male_reference = factors(MALE)
    female_reference = factors(FEMALE)
\end{lstlisting}

\section{Experiments and results}\label{sec:experiments}

This section tests the proposed \ac{DL}  method and architecture on two representative and public datasets. Five consolidated feature selection baselines are used: logistic, entropy, variance and PCA. The classification accuracy score has been selected to measure the quality of the results. Most of the experiments compare the results obtained by choosing different numbers of selected features. Four classification models are used in the experiments: neural networks, logistic regression, SVM and random forest. Cross validation has been implemented by using a 70\% training set, a 10\% validation set and a 20\% testing set. The chosen datasets to make the experiments are the popular MovieLens and the MyAnimeList. Both of them contain demographic information. MyAnimeList contains more than five million of ratings, and the selected MovieLens version has only 100,000 ratings; in this way we will test the proposed method on two datasets with very different sizes. Some relevant dataset facts are shown in \cref{tab:table-datasets}. We will show all the results from the MovieLens dataset, and the more representative ones from the MyAnimeList dataset. Two demographic features have been tested: gender and age; we have found little differences in their results. To maintain the paper in a reasonable size, and to avoid including redundant information, figures in this section are restricted to the gender results.

\begin{table}[!h]
    \centering
    \begin{tabular}{c|c|c|c|c}
         & \#users & \#items & \#ratings & scores \\\hline
        MovieLens & 943 & 1682 & 100,000 & 1 to 5 \\
        MyAnimeList & 69,600 & 9,927 & 5,788,207 & 1 to 10\\
    \end{tabular}
    \caption{Datasets used in the experimentation}
    \label{tab:table-datasets}
\end{table}

\begin{figure*}[!h]
    \centering
    \includegraphics[width=0.9\textwidth]{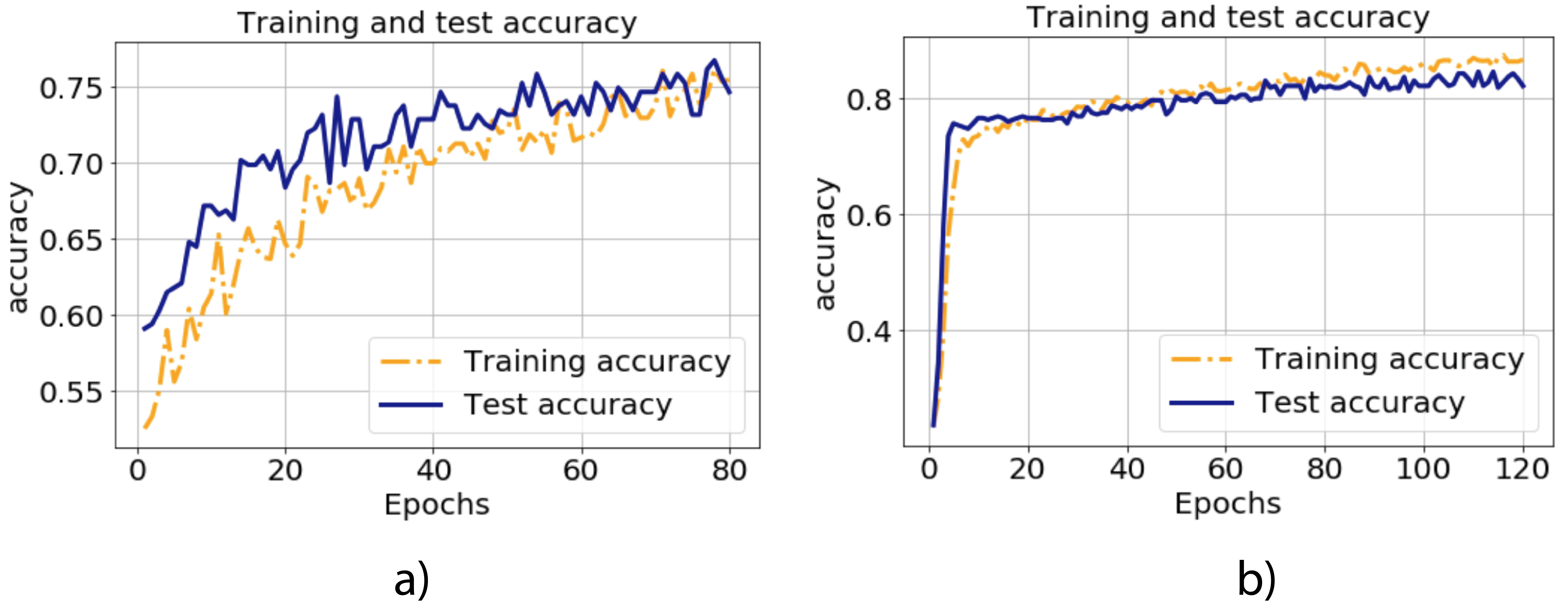}
    \caption{Training and test classification accuracy reached in the \ac{MLN} of the \DeepUnHide architecture. Gender classification (a), age classification (b). MovieLens dataset.}\label{fig:9}
\end{figure*}

Throughout the performed experiments, \ac{MF} has been processed by using a large number of factors: $K=100$; it makes possible to spread features among them. Once the \ac{MF} has been run on both datasets, the following step is to train the \ac{MLN} included in the \DeepUnHide Architecture (\Cref{fig:5}). We have classified users for both the gender and the age demographic features. The age groups are under 40 years old and 40 or more years old. \Cref{fig:9} shows the classification accuracy obtained in the MovieLens dataset for both the gender and the age demographic features. Regarding the MyAnimeList dataset, it reaches a 0.82 gender accuracy. The designed \ac{MLN} for MovieLens contains a 100 neurons input layer, a 10 neurons hidden layer (ReLU activation), a 0.3 dropout layer and finally the 2 neurons output layer to encode gender and age, using sigmoid activation. The chosen loss function is binary cross-entropy, and the optimizer is RMSprop. In the case of the MyAnimeList dataset the used \ac{MLN} is similar to the MovieLens one, with the only difference that the hidden layer contains 20 neurons.

\begin{figure*}[!h]
    \centering
    \includegraphics[width=0.85\textwidth]{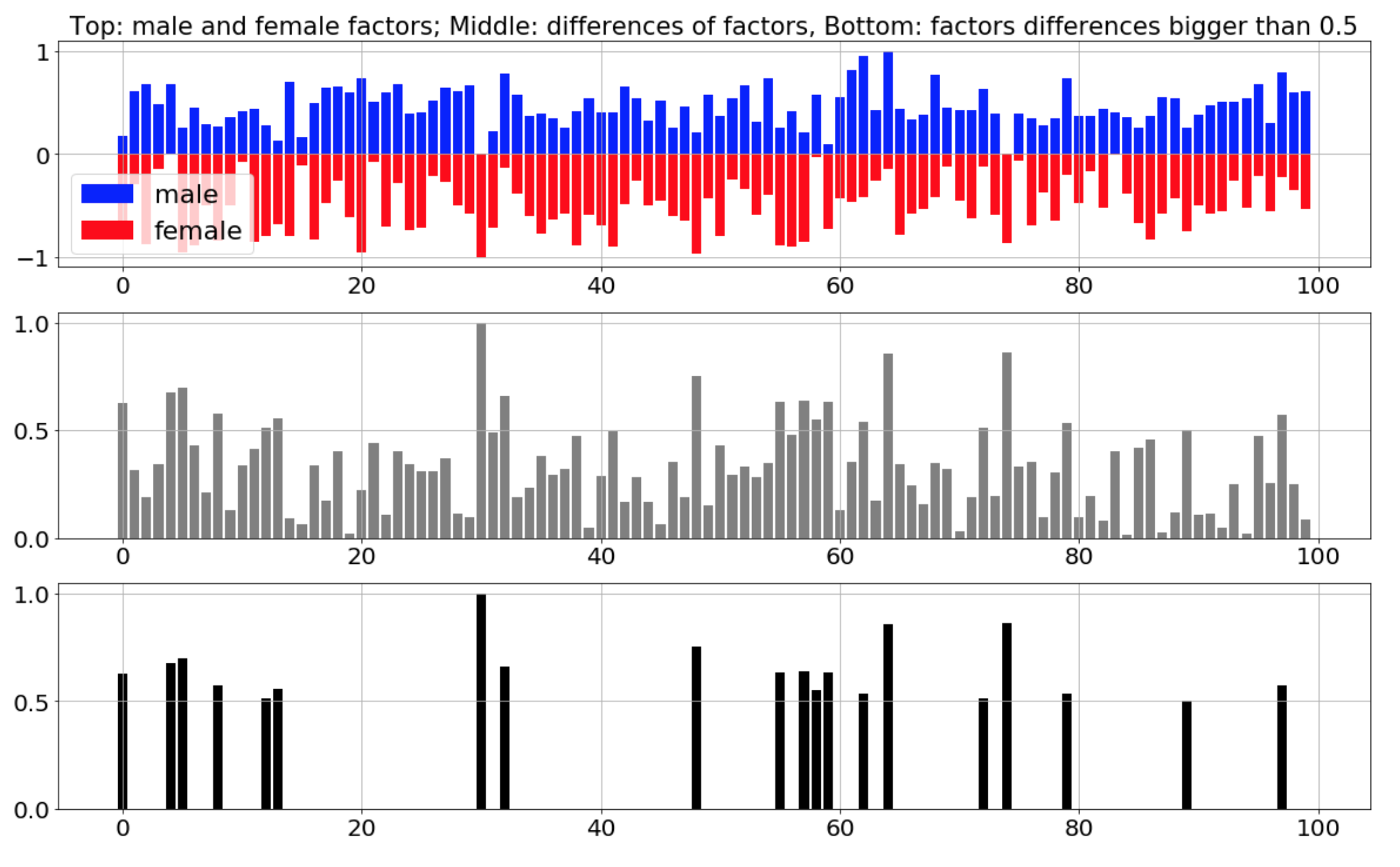}
    \caption{Male and female proportions encoded in each of the \ac{MF} factors; MovieLens dataset. $x$-axis: factor number; top graph: male and female proportions; middle graph: normalized absolute difference between male and female proportions; bottom graph: more relevant factors to the gender demographic feature}\label{fig:10}
\end{figure*}

Once the \DeepUnHide architecture \ac{MLN} (\Cref{fig:5}) has been trained, we run \DeepUnHide (\Cref{fig:6}) to obtain the demographic proportions of each user factor, as seen in \Cref{fig:2}. MovieLens gender (male, female) results are shown in \cref{fig:10}: its top graph draws the male (blue) and the female (red) proportions that each factor encodes ($x$-axis: factors). Representative factors are those which mostly encode male or mostly encode female: it helps to make a feature selection. Middle graph in \cref{fig:10} shows the normalized absolute difference of the female and male proportions: the largest the absolute difference, the better the factor distinguishes the feature. From these values we can select those whose normalized absolute difference exceed a threshold, obtaining the most relevant factors (selected features). 

\begin{figure*}[!h]
    \centering
    \includegraphics[width=0.87\textwidth]{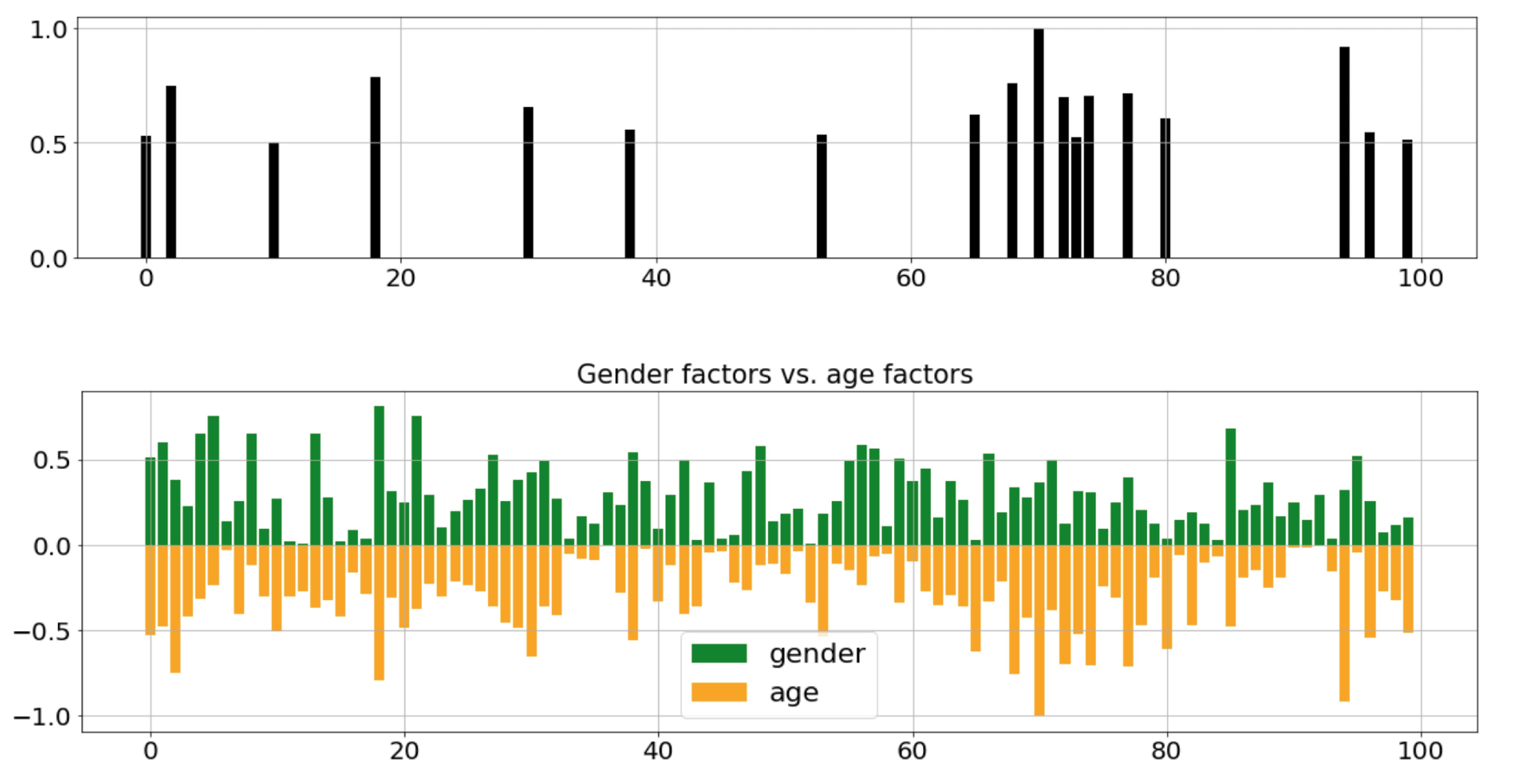}
    \caption{Top graph: more relevant factors to the age (young, senior) demographic feature; bottom graph: gender versus age proportions for each \ac{MF} factor. Movielens dataset. $x$-axis: factor number}\label{fig:11}
\end{figure*}

As an example, graph in the bottom of \cref{fig:10} shows the factors that overtake the 0.5 threshold: they are the result of the proposed \DeepUnHide feature selection. Making the same process to the age (young, senior) demographic feature, in the MovieLens dataset we obtain the results shown in the top graph of \cref{fig:11}. Please note that the same factor can be relevant to two different demographic features (such as factor 0 in \cref{fig:10} and \cref{fig:11}), although it will not be the usual situation when the number of the \ac{MF} factors is high. Bottom graph in \cref{fig:11} confronts the proportions of the demographic gender and age features for each \ac{MF} factor. Explanation of recommendations can be done using this information: since we know each demographic importance for each hidden factor, it is possible to extract this information from the prediction dot product, as shown in \cref{fig:2}.

Please note that the bottom graph in \cref{fig:10} shows the set of factors that best discriminate the gender demographic feature, whereas the top graph in \cref{fig:11} shows the set of factors that best discriminate the age demographic feature. All these factors have been obtained by using a threshold value. Another approach is to select the $N$ factors that best discriminate the desired demographic feature: instead of using the indicated threshold, we just take the $N$ most promising factors. The first two experiments in this section compare the classification quality results obtained by using different $N$ values (different number of factors). These are feature selection experiments; it is expected that the more the $N$ value, the best the quality results. It is also expected that a reduced number of factors can provide accurate classification values. Finally, the proposed \DeepUnHide method should show better scores and trends than the baselines do. The first experiment makes use of an $N$ range from 1 to 20; the second experiment uses an $N$ range from 5 to 70; finally, the third experiments fixes $N$ to 50.

\sloppypar{Once the \DeepUnHide feature selection is made, we have designed three experiments to test that it is correct and that it improves the state of the art. The three experiments test the classification accuracy quality measure, and all of them compare the proposed approach with several state of art feature selection methods: logistic~\cite{Ng2020}, entropy~\cite{Jiang2015Jul}, variance~\cite{Wang2018Aug} and PCA~\cite{Jolliffe2002}. Additionally, a random baseline is used. Each of the three experiments is explained in a separated subsection. The first experiment is focused on the four generated users: the representative male, female, young and senior; it uses an \ac{MLN} to classify these four vectors of factors. The second experiment classifies all the datasets users by means of an \ac{MLN}. Finally, the third experiment classifies all the datasets users by means of several \ac{ML} classification models.}

\subsection{Classification of the representative users applying a neural network}

\begin{figure*}[!h]
    \centering
    \includegraphics[width=0.7\textwidth]{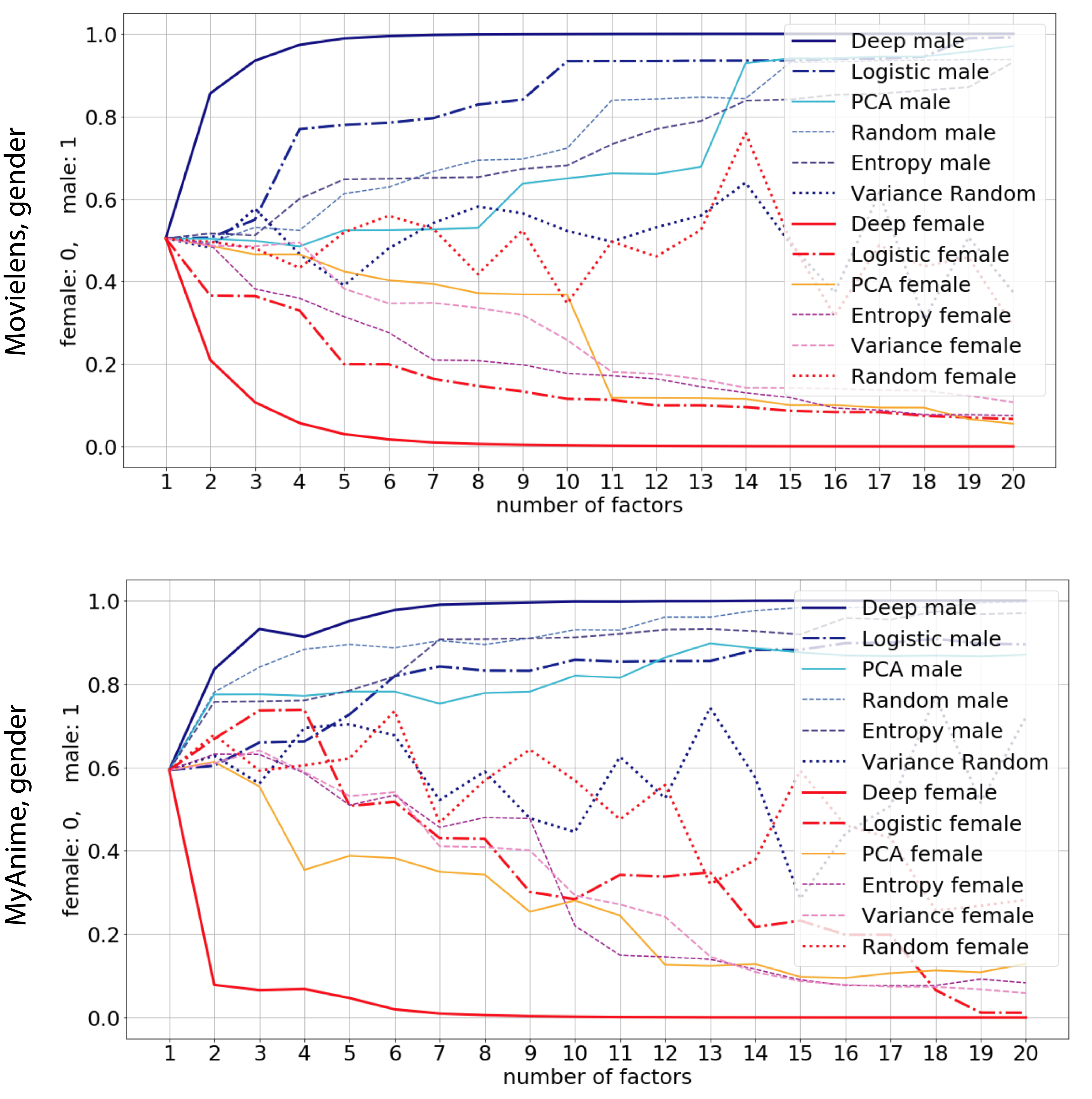}
    \caption{Classification results of the proposed \DeepUnHide  method (Deep for short) compared to the feature selection baselines: logistic~\cite{Ng2020}, entropy~\cite{Jiang2015Jul}, variance~\cite{Wang2018Aug} and PCA~\cite{Jolliffe2002}. $x$-axis: number of selected factors, $y$-axis: correct classification values (0 for female and 1 for male)}\label{fig:12}
\end{figure*}

This experiment uses the factors of the representative male and female (\Cref{fig:8}, iteration $n$). The hypothesis is that correct classification can be achieved by using a reduced set of the selected features. Classification is performed running forward (predicting) the same \ac{MLN} of the \DeepUnHide architecture (\Cref{fig:8}). Predictions near to the value 1 can be considered users classified as ‘male’, whereas predictions near to the value 0 can be considered as ‘female’ users (same with ‘young’ and ‘senior’). Different classification processes are made using different numbers of factors (from 1 to 20). \Cref{fig:12} shows the obtained results for the tested datasets: MovieLens (top graph) and MyAnimeList (bottom graph) when the gender feature is chosen. The male (blue color) correct classification value is the number 1 ($y$-axis) and the female (red color) correct classification value is the number 0 ($y$-axis). Both graphs in \cref{fig:12} show, as expected, that increasing the number of selected factors the classification accuracy raises. Solid lines in \cref{fig:12} correspond to the proposed \DeepUnHide method (``Deep'', for short in legends). It can be seen that our method works fine even with a very reduced number of selected factors: using just 3 factors it can correctly classify with small errors. All the baselines need a larger number of factors than the proposed method to reach a similar accuracy; logistic is the best baseline on MovieLens, whereas entropy is better in MyAnimeList. None of them can compete with \DeepUnHide. Please note that experiments that selects a very low number of factors (one to three factors) can return an ambiguous classification result. They do not manage enough information to correctly classify users; e.g.: the 0.5 and 0.6 classification values in \cref{fig:12}, when only one factor is selected. The demographic age classification results are very similar to the gender ones.

\subsection{Classification of all the users applying a neural network}

\begin{figure*}[!h]
    \centering
    \includegraphics[width=0.95\textwidth]{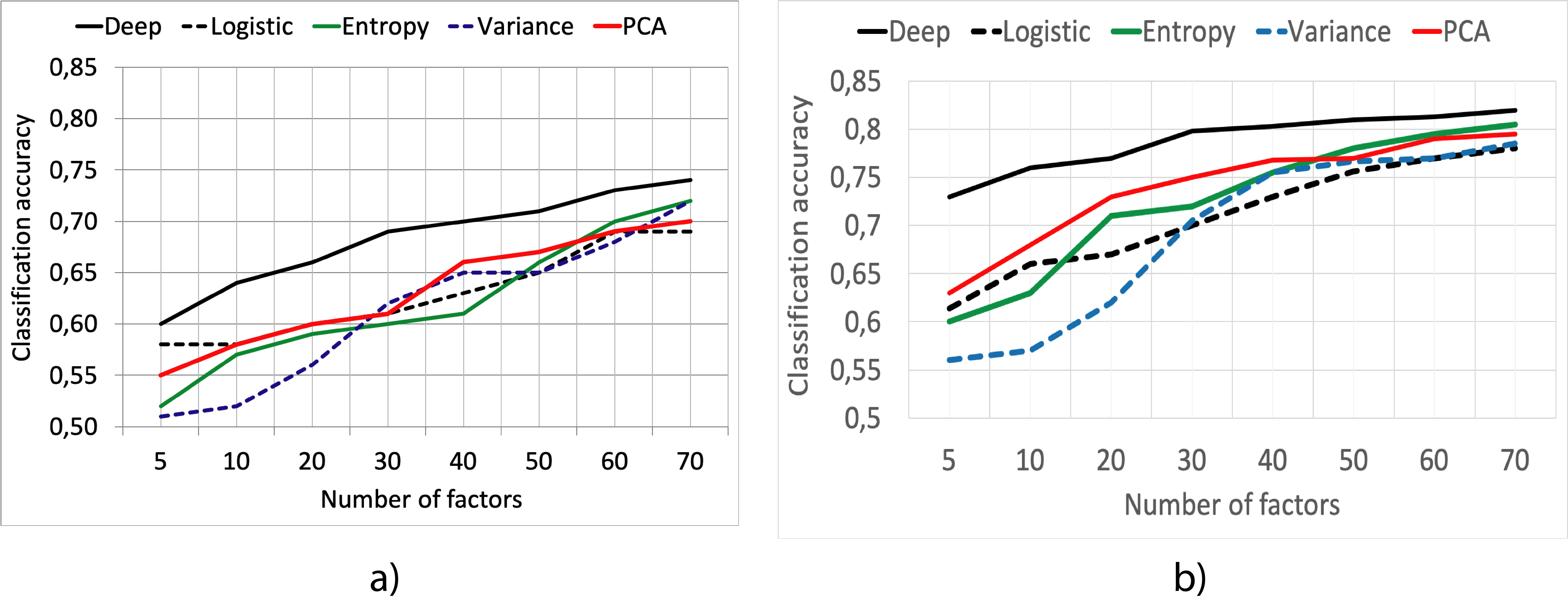}
    \caption{Classification accuracy of the users based on their gender; a) Movielens, b) MyAnime. $x$-axis: number of selected factors, $y$-axis: accuracy. Proposed method: \DeepUnHide (Deep for short). Baselines: logistic~\cite{Ng2020}, entropy~\cite{Jiang2015Jul}, variance~\cite{Wang2018Aug} and PCA~\cite{Jolliffe2002}}\label{fig:13}
\end{figure*}

The previous experiment did not classify all the testing users in the dataset. It just classified the representative male user and the representative female user (by using different numbers of selected factors). Experiments in this section make use of an \ac{MLN} to classify all the users attending to their gender demographic feature. Hypothesis here is the same than in the previous experiment one: correct classification can be achieved by using a reduced set of selected features. Experiments in this section have been performed varying the number of selected factors. \Cref{fig:13} shows the obtained results: as expected, accuracy increases when the number of selected factors grows. The MyAnimeList dataset reaches better classification results; \DeepUnHide (Deep for short) improves all the baselines accuracy for all the tested number of selected factors.

\subsection{Classification of all the users applying several machine learning models}

\begin{figure*}[!h]
    \centering
    \includegraphics[width=0.95\textwidth]{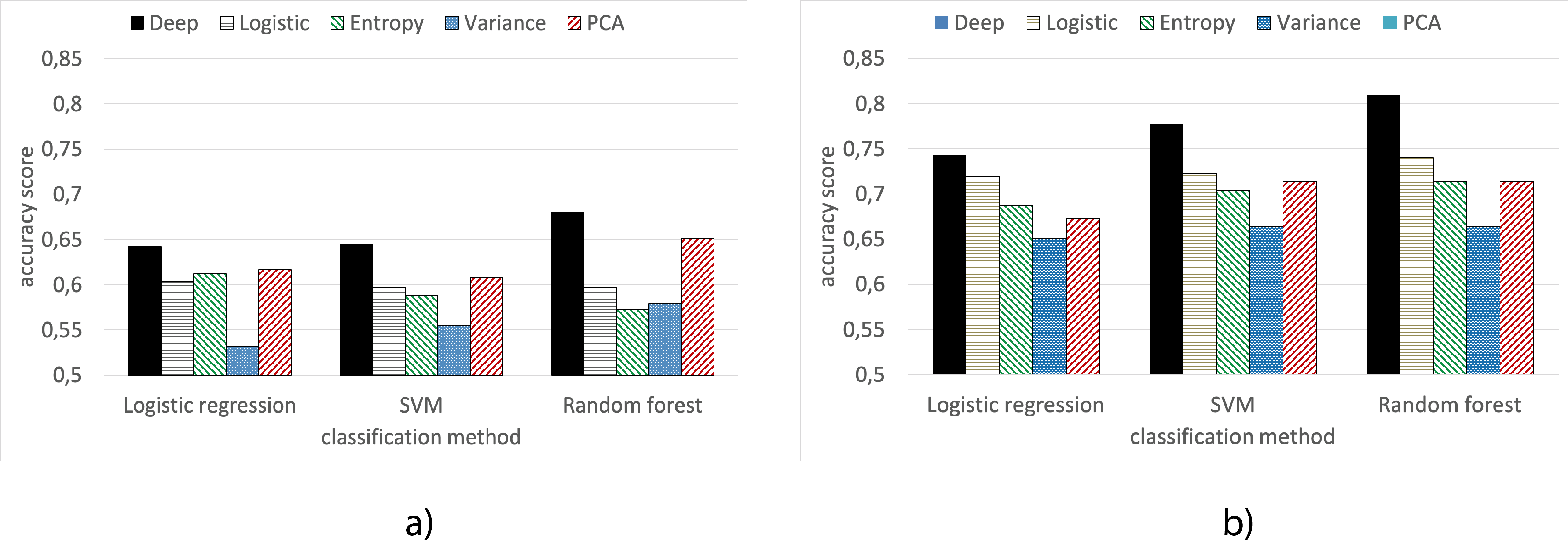}
    \caption{Classification accuracy of the users; a) Movielens, b) MyAnime. $x$-axis: classification model, $y$-axis: accuracy. Proposed feature selection: \DeepUnHide (Deep for short). Baselines: logistic~\cite{Ng2020}, entropy~\cite{Jiang2015Jul}, variance~\cite{Wang2018Aug} and PCA~\cite{Jolliffe2002}. Demographic feature: gender}\label{fig:14}
\end{figure*}

Since the proposed feature selection method and its architecture are based on the \ac{DL} model, and in the previous section it was tested the accuracy by means of a \ac{DL}  classifier, it has been considered convenient to make some quality testing based on different classification models. In particular, the \ac{ML} logistic regression, SVM and random forest have been chosen. In this section we compare the accuracy score obtained using the mentioned models and applying them to both the proposed \DeepUnHide method and the selected baselines. \Cref{fig:14} shows the results obtained in the: a) MovieLens dataset and b) MyAnimeList dataset. The number of selected factors has been fixed to 50 (half of the whole available factors).  In the same line that previous experiments, the proposed Deep feature selection gets better accuracy than the baselines in all the cases. MyAnimeList reaches better classification accuracy than MovieLens, and random forest is the \ac{ML} model with better results, although they are worse than the \ac{DL}  model (comparing \cref{fig:13,fig:14}).

\subsection{Unhiding the hidden factors in the MovieLens dataset}

\DeepUnHide has been designed to facilitate the understanding of hidden factors in a \ac{MF} based \ac{RS} by extracting demographic information from users. This information can be used to explain the recommendations performed by the \ac{RS}. It can be done by making use of the hidden factors of the majority archetypal user $p_M$ and the minority archetypal user $p_m$ (see~\cref{sec:math-formulation}). By selecting their highest hidden factors, we can obtain the most representative factors of the majority group, $K_M$, and the minority group, $K_m$.

In this experiment we define an index to assign a minority value to the $i$-th item (say, femininity), $w_m(i)$, and equivalently an other index to assign a majority value to each item (say, masculinity), $w_M(i)$. These two weights can be obtained by comparing the archetypal users $\tilde{p}_m$ and $\tilde{p}_M$ with the hidden factors of the item $i$, $q_i$. To be precise, we define the minority weight $w_m(i)$ of each item $i$ as
$$
w_m(i) = \sum_{j \in K_m} \tilde{p}_{m}^j \cdot q_{i}^j.
$$

Analogously, we define a majority weight $w_M(i)$ of each item $i$ as
$$
w_M(i) = \sum_{j \in K_M} \tilde{p}_{M}^j \cdot q_{i}^j.
$$

We focus on the MovieLens dataset, by assigning \textit{male} users to the majority group and \textit{female} users to the minority group. Using the previously computed archetypal male user, $\tilde{p}_M$, and the archetypal female user, $\tilde{p}_m$, we compute the coefficients $w_M(i)$ and $w_m(i)$ for each item of the dataset. The 10 most representative movies of the minority group (i.e. the top 10 items with the highest $w_m(i)$) are the following: 

\begin{enumerate}
    \item Evita (1996)
    \item Crucible, The (1996)
    \item Dirty Dancing (1987)
    \item Nell (1994)
    \item Rosewood (1997)
    \item Dante's Peak (1997)
    \item Jungle2Jungle (1997)
    \item On Golden Pond (1981)
    \item My Best Friend's Wedding (1997)
    \item Little Women (1994)
\end{enumerate}

On the other hand, the following 10 movies as the most representative of the majority group (i.e.\ the top 10 items with the highest $w_M(i)$): 

\begin{enumerate}
    \item Fifth Element, The (1997)
    \item Trainspotting (1996)
    \item Crumb (1994)
    \item Die Hard (1988)
    \item Clerks (1994)
    \item Aliens (1986)
    \item Miller's Crossing (1990)
    \item Lost Highway (1997)
    \item Brazil (1985)
    \item Dances with Wolves (1990)
\end{enumerate}

For this experiment we have used the top $N=20$ hidden factors of the archetypal users to compute $K_M$ and $K_m$ and all movies with less than 75 ratings has been filtered out to avoid cold start situations.

\section{Conclusions}\label{sec:conclusions}

An innovative approach to unhide demographic features in matrix factorization is presented. It uses the gradient based localization concept, borrowed from the deep learning image processing. The obtained representative user vector for each demographic feature (say ‘gender) serves to make the feature selection. Results show an important improvement in the classification accuracy score when the selected features are applied, compared to the baseline methods. Thus, we can assure that the proposed deep learning method and architecture accurately catch the hidden semantic of the matrix factorization factors. The obtained results open the door to reach improvements on several representative research fields in the recommender systems area. Recommendation explanation can be addressed by translating from the obtained demographic information to a visual representation of demographic features. Fairness is another important research field where the proposed method has a direct application: fair recommendations can be made by weighting those factors that belong to the biased group of users. The proposed method can also be applied to recommendation of groups of users, making use of the gradient-obtained representative user that can act as virtual user for the group.

\bibliographystyle{abbrv}
\bibliography{DeepUnHide.bib}

\end{document}